# Stochastic Model Based Proxy Servers Architecture for VoD to Achieve Reduced Client Waiting Time

Dr T R GopalaKrishnan Nair[1] and M Dakshayini[2],

[1]Director, Research and Industry Incubation Centre, DSI,
Bangalore, India

[2] Research Scholar, Dr. MGR University. Working with Dept. of ISE, BMSCE, Bangalore.
Member, Multimedia Research Group, Research Centre, DSI,
Bangalore, India

**Abstract**
In a video on demand system, the main video repository may be far away from the user and generally has limited streaming capacities. Since a high quality video's size is huge, it requires high bandwidth for streaming over the internet. In order to achieve a higher video hit ratio, reduced client waiting time, distributed server's architecture can be used, in which multiple local servers are placed close to clients and, based on their regional demands video contents are cached dynamically from the main server. As the cost of proxy server is decreasing and demand for reduced waiting time is increasing day by day, newer architectures are explored, innovative schemes are arrived at. In this paper we present novel 3 layer architecture, includes main multimedia server, a Tracker and Proxy servers. This architecture targets to optimize the client waiting time. We also propose an efficient prefix caching and load sharing algorithm at the proxy server to allocate the cache according to regional popularity of the video. The simulation results demonstrate that it achieves significantly lower client's waiting time, when compared to the other existing algorithms.

## 1. Introduction

The tremendous growth of World Wide Web has resulted in an increase of bandwidth consumption throughout the internet. Proxy caching has been recognized as an effective technique to reduce network traffic. Caching is also an important mechanism for improving both the performance and operational cost of multimedia networks [10,13]. Recent web video access patterns show frequent requests for a small number of popular objects at popular sites. So a popular video can be streamed to the same network link once per request. In the absence of caching, this approach results in server over load, network congestion, higher request-service delay, and even the higher possibility of rejection of a clients request. Caching the partial or the complete videos which has a high demand locally at the proxy servers solves all these problems. This reduces the main server load by distributing the load across the network [3].

*VoD* system usually has several servers and distributed clients over the entire network. These servers contain prerecorded videos and are streamed to the clients upon request from the clients. Proxy cache attempt to improve performance of the overall network communication in three ways [9]:
i Reduce the request-service delay associated with obtaining documents (because the proxy cache is placed typically closer to the user).
ii. Lower the network traffic (the documents served already are available to the user for next time so less load on the network)
iii. Reduce the Network cost.
In recent years, to reduce the request-service delay and bandwidth demand between the Main multimedia server and the proxy servers, a number of caching and buffering techniques have been proposed. Most of these techniques use proxy servers with large storage space for caching videos which are requested frequently. The cached data is used to serve the future requests and only the un cached portions of the video are downloaded from the Main servers [2, 12].
Proxy servers have been widely used for multimedia contents to decrease the startup delay and to reduce the load of the Main multimedia server. Recent works investigate the advantages of connected proxy servers within the same intranet [3, 4 and 8].

## 2. Related work

This section briefly discusses the previous work as follows, Tay and pang have proposed an algorithm in Ref.[3] called *GWQ* (Global waiting queue) which reduces the initial startup delay by sharing the videos in a distributed loosely coupled *VoD* system by balancing the load between the lightly loaded proxy servers and heavily loaded proxy servers in a distributed *VoD*. So whenever the local server is busy, the request will be serviced from the remote server. This introduces the additional network traffic that flows from remote servers. They have replicated the videos evenly in all the servers, for which the storage capacity of individual proxy server should be very large to store all the videos. This may not allow each server to store replicas of more number of videos. Our proposed scheme replicates only regionally (local and global) popular videos using dynamic buffer allocation algorithm[2] there by utilizing the proxy server storage space more efficiently to store replicas of more number of videos. In [4] Sonia Gonzalez, Navarro, Zapata proposed a more realistic partial replication and load sharing algorithm PRLS to distribute the load in a distributed *VoD* system. In their research, they have demonstrated that their algorithm maintains a small initial start up delay using less storage capacity servers by allowing partial replication of the videos. They store the locally requested videos in each server. Our work differs by caching the initial some portion of the video as prefix-1 at proxy and next part of the video as prefix-2 at tracker based on local and global popularity using dynamic buffer allocation algorithm [2]. S.-H. Gary Chan, Fouad Tobagi in [7] considers the exchange of cached contents with the neighboring proxy server without any coordinator. Our approach differs, in which we have made a group of proxy servers with a coordinator (Tracker) to make the sharing of videos more efficient. Another approach to reduce the aggregated transmission cost has been discussed in [6] by caching the prefix and prefix of suffix at proxy and client respectively. Since the clients are not trustable, and can fail or may leave the network at any time without any notice, they have adopted an additional mechanism to verify the client and cached data at client, which increases the overhead of such verification. Both searching of the video in the whole cluster of proxy servers, and the verification process increases the client's waiting time.

So in order to minimize the client waiting time and network traffic in the VoD system, in this paper, we present a novel 3 layer architecture of distributed proxy servers, for serving videos with a target to optimize the client waiting time. This architecture consists of a *Main multimedia server* [*MMS*], which is very far away from the user and is connected to a set of *trackers* [*TR*]. Each tracker is in turn connected to a group of proxy servers [PSs] and these proxy servers are assumed to be interconnected in a ring pattern, this arrangement of cluster of proxy servers is called as *Local Proxy servers Group[LPSG($L_p$)]*. Each of such *LPSG*, which is connected to *MMS*, is in turn connected to its left and right neighboring *LPSG* in a ring fashion through its tracker. We also propose an efficient regional popularity based prefix caching and load sharing algorithm (RPPCL). This algorithm efficiently allocates the cache blocks to the video according to their local popularity and also shares the videos present among the PSs of the LPSG. Hence our approach increases the video hit rate and reduces the client waiting time, network usage on MMS to PS path.

The main aim of arranging the group of proxy servers in the form of *LPSG* is to *provide* the following advantages.

- *Reduced Client waiting time*: replicating the videos at *PS*s of $L_p$ based on their local popularity, and sharing of these videos among the *PS*s of $L_p$ can provide the service to the clients immediately as they request.
- *Increased aggregate storage space*: by distributing large number of videos across the *PS*s and *TR* of $L_p$, high cache hit rate can be achieved. For example, if 10 *PS*s within a *LPSG* managed 500 Mbytes each, total space available is 5 GB. 200 proxies of *LPSG* could store about 100 GB of movies.
- *Load reduction*: replication of the videos among the *PS*s of $L_p$ based on their regional popularity, allows more number of clients to get serviced from $L_p$. This reduces the communication with the main multimedia server and in turn its load.
- *Scalability*: by adding more number of PSs the capacity of the system can be expanded. Interconnected *TR*s increases the system throughput

The organization of rest of the paper is as follows: In section 3 we present a Model of the problem, Section 4 describes the proposed approach and algorithm in detail, In section 5 we present a simulation model, Section 6 presents the simulation results and comparison of RPPCL, GWQ and PRLS algorithms, Finally, in section 7 we conclude the paper and refer to further work.

## 3. Stochastic Model of the Problem

Let *N* be a stochastic variable representing the group of videos. It may take the different values for (videos) $V_i$ (i=1,2 . . N) and the probability of the video $V_i$ being asked is $p(V_i)$. Let the set of values $p(V_i)$ be the probability mass function. Since the variable must take one of the values, it follows that $\sum_{i=1}^{N} p(v_i) = 1$. So the estimation of the probability of requesting $V_i$ video, is

$$p(V_i) = \frac{n_i}{I}.$$

Where $I$ is the total number of observations and $n_i$ is the number of requests for $i^{th}$ video. A cumulative distribution function denoted as $P(V_i)$ is the function that gives the probability of a request (random variable's) being less than or equal to a given maximum value.

We assume that client's requests (X/hr) arrive according to Poisson process with λ as shown in Fig.2 of simulation model. Let $S_i$ be the size (duration in minutes) of $i^{th}$ video with mean arrival rates $\lambda_1 \ldots \lambda_N$ respectively that are being streamed to the users using $M$ proxy servers (*PSs*) of $J$ LPSGs ($L_p$ p=1..J).

Each *TR* and *PSq(q=1..M)*, has a caching buffer large enough to cache total $P$ and $B$ minutes of $H$ and $K$ number of videos respectively.

i.e. $p = \sum_{i=1}^{H}(pref-2)_i$ and $B = \sum_{i=1}^{K}(pref-1)_i$

Every $i^{th}$ video $V_i$ is divided into 3 parts, first $W_1$ minutes of each video $V_i$ is referred to as prefix-1 $(pref-1)_i$ of $V_i$. If $V_i$ is globally popular then it is replicated at all M PSs otherwise it is replicated across $L$ PSs of $L_p(p=1..J)$, in which the frequency of accessing the video $V_i$ is high. Next $W_2$ minutes of video $V_i$ is referred to as prefix-2 $(pref-2)_i$ of $V_i$ is cached at *TR* of $L_p$ and the rest of the video is referred to as suffix of the video and is stored at *MMS* as shown in fig.1. This arrangement of replicating the popularity based (pref-1) at $L$ *PSs*, helps the system to serve the request immediately as the request arrives. It also keeps the queue length QL very small.

| W1 min.(Vi) | W2 min(Vi) | |
|---|---|---|
| Prefix-1 | Prefix-2 | Suffix S(Vi) |

Fig. 1 Parts of the video Vi

Depending on the probability of occurrence of user requests to any video, the popularity and size of (*pref-1*) and (*pref-2*) of the videos to be cached at *PS* and *TR* respectively are determined. That is size *(W)* of (*pref-1*) and (*pref-2*) for $i^{th}$ video is determined as.
i.e.

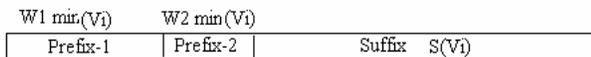

So $W(pref-1)_i = x_i \times S_i$ where $0 < x_i < 1$
$W(pref-2)_i = x_i \times (S_i - (pref-1)_i)$
where $0 < x_i < 1$

Where $x_i$ is the probability of arrival of requests for the $i^{th}$ video from last t minutes, and $n_i$ is the total number of requests for video $V_i$. Let $b_i$ be the available bandwidth for $V_i$ between the proxy server and Main multimedia Server. After requesting for a video $V_i$ at $PS_q$, the streaming of that video $V_i$ may be delayed by

$$Wt_i^{PSq} = T(pref-1)_i^{PSq} \text{ where } i=1..N, q=1...M$$

Where $T$ is the time required to retrieve and initiate the streaming of $(pref-1)_i$ from $PS$ to the requested user(ps-user). Subsequently by the end of $w_1$ minutes $(pref-2)_i$ will be streamed from $TR$ to user through $PS$ ((TR-PS)(PS-user)). By the end of $w_2$ minutes, $(S-(pref-1)-(pref-2))_{Vi}$ will be streamed from *MMS* to user through *PS* ((MMs-TR)(TR-PS)(PS-user)) in continuous to $(pref-1)_i$.

Another output stochastic variable $y$ is the average waiting time for all the clients. Thus $y$ is a sample mean of client delays $Wt_1, Wt_2...Wt_N$.

That is $y = \frac{1}{Q} \sum_{i=1}^{Q}(Wt_i)$

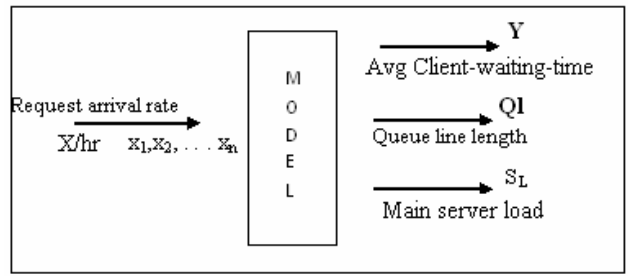

Figure. 2 Simulation Model

Let $Q$ be another stochastic variable represents the number of requests served immediately from $PS_q$, and $Wt(\ )$, is the non-linear function. The optimization problem is to maximize the number of clients $Q$ served *from* $PS_q$ immediately, by replicating the popularity based *(pref-1)* videos at $L$ *PSs* using dynamic buffer allocation algorithm [2]. Also to minimize the average user waiting time $y$ at *PS* by sharing the videos cached among the *PSs* of $L_p$. This can be formulated as follows:

Minimize *Wtime* is $y = \frac{1}{Q} \sum_{i=1}^{Q}(Wt_i)$

Subject to

$B = \sum_{i=1}^{K}(pref-1)_i$, $P = \sum_{i=1}^{H}(pref-2)_i$

$(pref-1) > 0$ and $(pref-2) > 0$

## 4. Proposed Architecture and Algorithm

4.1. Overview of the proposed Architecture

The proposed 3 layer architecture is as shown in Fig.4.
This architecture consists of a *MMS*, which is connected to a group of trackers (*TRs*), Each *TR* has various modules. As shown in the fig. 3, they are
1. *Interaction Module ($IM_{TR}$)* – Interacts with the *PS* and *MMS*.
2. *Service Manager ($SM_{TR}$)* – Handles the requests from the *PS*.
3. *Database* – Stores the complete details of presence and size of *(pref-1)* of videos at all the *PSs*.
4. *Video distributing Manager(VDM)* – Responsible for deciding the videos, and sizes of *(pref-1), (pref-2)* of videos to be cached. Also handles the distribution and management of these videos to group of *PSs,* based on video's global and local popularity.

Each *TR* is in turn connected to a set of *PSs*. These *PSs* are connected among themselves in a ring fashion. Each *PS* also has various modules such as,
1. *Interaction Module ($IM_{PS}$)* – Interacts with the user and *TR*.
2. *Service Manager ($SM_{PS}$)* – Handles the requests from the user,
3. *Popularity agent (PA)* – Observes and updates the popularity of videos at PS as well as at *TR*,
4. *Cache Allocator (CA)* – Allocates the Cache blocks using dynamic buffer allocation algorithm [2]. Also to each of these proxy servers a large number of users are connected [*LPSG]*. Each proxy server is called as a parent proxy server to its clients. All these *LPSGs* are interconnected through their *TR* in a ring pattern as shown in fig. 4.

The *PS* caches the *(pref-1)* of videos distributed by *VDM,* and streams this cached portion of the videos to the clients upon the request through *LAN* using its less expensive bandwidth.
We assume that,
1. The *TR* is also a *PS* with high computational power and large storage compared to other proxy servers, to which clients are connected. It has various modules, using which it coordinates and maintains a database that contains the information of the presence of videos, and also size of *(pref-1)* and *(pref-2)* of video in each *PS* and *TR* respectively
2. Proxies and their clients are closely located with relatively low communication cost[1]. The Main server in which all the videos completely stored is placed far away from *LPSG,* which involves high cost remote communication.

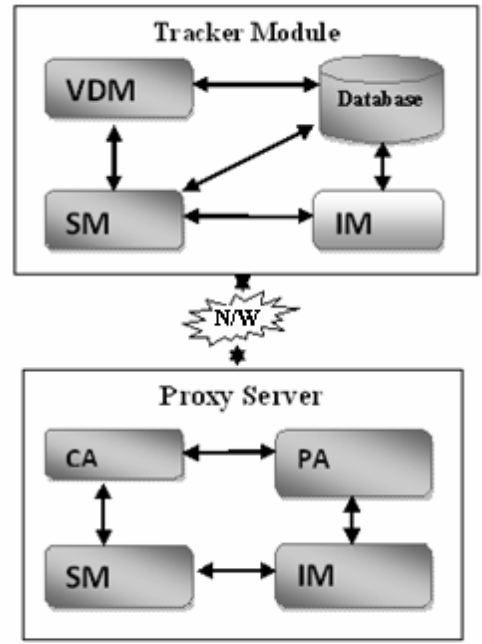

Figure 3  Modules of Proxy server and Tracker

3. The *MMS*, the *TR* and the *PSs* of *LPSG* are assumed to be interconnected through high capacity optic fiber cables. In the beginning, all the $N_v$ videos are stored in the *MMS*. The distribution of the selected *N* of $N_v$ videos among *M* *PSs* of the *LPSG* is  done by *VDM* as follows. First, all the *N* videos are arranged with respect to their popularity at jth *LPSG* . The popularity of a video is   defined as the probability of frequency of requests to this video per threshold time t. Here, we   assume that the frequency of requests to a video follows Zipf law of distribution. The video distribution module divides N videos    into two subgroups- the globally popular $k(0 <= k <= N)$ videos like Cartoons, and locally popular $N - k$ videos –such that former small subgroup is replicated in all the *PSs* and the later subgroup is cached at *PS* of  $L_p$ based on the  local demand for the videos 4.2. Proposed Algorithm
 Since the storage cache space of both *PS* ($C_{PSq}$ )and *TR* ($C_{TR}$ ) is limited, the VDM of the *TR* first executes the decision making algorithm to fix up the  sizes(segments) of *(pref-1)* and *(pref-2)* of videos to be cached at  $C_{PSq}$ and in its cache $C_{TR}$ respectively. Then caching is done using dynamic buffer allocation algorithm [2]. The corresponding entry is updated in its database at *TR*. Whenever a client at $PS_q$ wishes to play a video $V_i$, it first sends a request to its parent proxy $PS_q$, the $SM_{PSq}$ immediately starts streaming the *(pref-1)* of video requested to the client, if it is present in its cache. So waiting time is almost negligible.  And informs the $SM_{TR}$

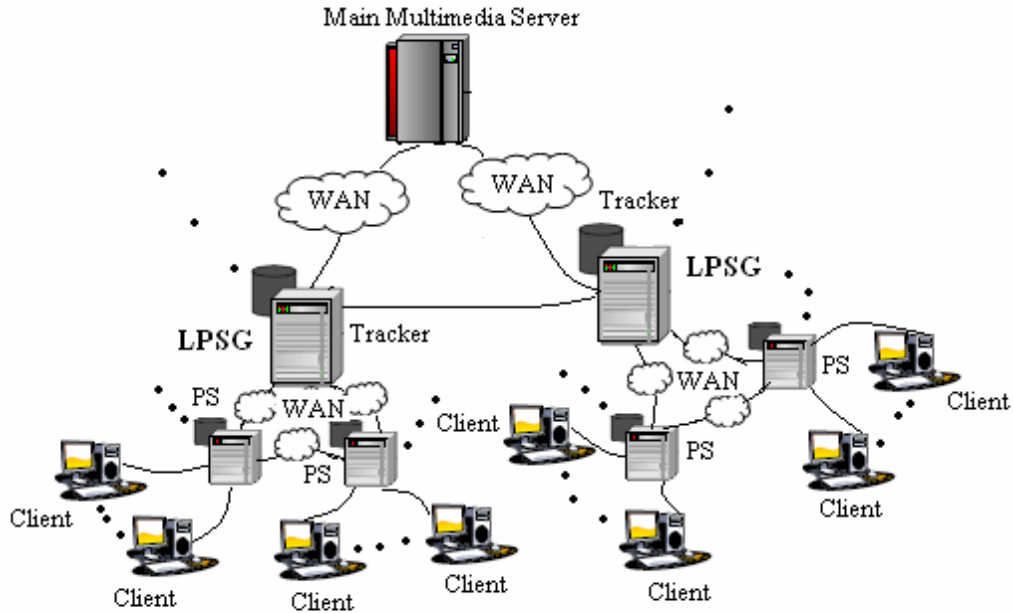

Figure 4  Proposed Architecture

**Proposed algorithm**

When there is a request for a video $v_i$ (at a particular proxy $PS_q$ of $L_p$, do the following:

If $(V_{req} \in PS_q)$

    *(pref-1)$_{Vreq}$* is streamed immediately to the user ( $y$ = time required to stream (pref-1) from proxy - user)

  So $wt_{Vreq} = wt(p\text{-}1)_{Vreq}^{p-u}$

else - pass the request to the TR(Lp)

    if $(V_{req} \in PS(Lp))$

        If *(PS(Lp)* is left or right *NBR(PS$_q$)*

            $SM_{TR}$ streams *(pref-1)$_{Vreq}$* from *NBR(ps$_q$)*, *(pref-2) $_{Vreq}$* from its cache and the remaining portion from *MMS*

            $wt_{Vreq} = wt(p\text{-}1)_{Vreq}^{(p-p)+(p-u)}$ ( $y$ = time required to stream (pref-1) from proxy- proxy & proxy - user)

        else

            $SM_{TR}$ streams the *(pref-1)$_{Vreq}$* from *OTR(PS$_q$)*, *(pref-2) $_{Vreq}$* from its cache and the remaining portion from *MMS* to-User thru *PS$_q$* using optimal path found

            $wt_{Vreq} = wt(p\text{-}1)_{Vreq}^{(p-p)+(p-u)}$ ( $y$ = time required to stream (pref-1) from proxy- proxy & proxy - user)

    else

       Pass the request to left or right *TR(NBR(Lp))*

         if *(V$_{req}$ $\in$ NBR(Lp))*

           *TR(NBR(Lp))* streams the $V_{req}$ from *NBR(Lp)*-user thru *TR(L$_p$)*

           $wt_{Vreq} = wt[(p\text{-}1) + (p\text{-}2)]_{Vreq}^{(t-t)+(t-p)+(p-u)}$ ( $y$ = time required to stream (pref-1) from tracker – Tracker , tracker – proxy & proxy - user)

        else

           *TR(Lp)* downloads the complete $V_{req}$ from *MMS* and streams to the user

           $wt_{Vreq} = wt(S)_{Vreq}^{(s-t)+(t-p)+(p-u)}$ ( $y$ = time required to stream (pref-1) from MMS -TR, TR-PS & PS-user)

         Also caches the *(pref-1) and (pref-2)* of $V_{req}$ at $PS_q$ using Dynamic Buffer allocation algorithm[ 2].

corresponding entry is updated in its database at *TR*. Whenever a client at $PS_q$ wishes to play a video $V_i$, it first sends a request to its parent proxy $PS_q$, the $SM_{PSq}$ immediately starts streaming the *(pref-1)* of video requested to the client, if it is present in its cache. So waiting time is almost negligeble. And informs the $SM_{TR}$ to initiate the streaming of *(pref-2)* of $V_i$, then the $IM_{TR}$ coordinates with *MMS* to download the remaining portion $(S-(pref-1)-(pref-2))_{Vi}$ of the video $V_i$.

If it is not present in its cache, the $IM_{PSq}$ forwards the request to its parent *TR*, *VDM* at *TR* searches its database using perfect hashing to see whether it is present in any of the *PSs* in that $L_p$. If the $V_i$ is present in any of the *PSs* in that $L_p$, then the *VDM* checks whether the PS in which the $V_i$ found is neighbor to the requested $PS_q$ *[NBR(PS_q)]*.

If so, the *VDM* intimates the same to $SM_{TR}$ which initiates the streaming of the $(pref-1)_{Vi}$ from that $NBR(PS_q)$, and $(pref-2)_{Vi}$ from its cache, to the requested $PS_q$ and the same is intimated to the requested $PS_q$. Then the $IM_{TR}$ coordinate with *MMS* to download the remaining portion $(S-(pref-1)-(pref-2))_{Vi}$, and hence the client waiting time is very small.

Otherwise, if it is not *[NBR(PS_q)]* and is present in more than one PS of $L_p$ then $SM_{TR}$ selects one PS such that, the path from selected PS to $PS_q$ should be optimum and initiates the streaming of the $(pref-1)_{Vi}$ from the selected PS, and $(pref-2)_{Vi}$ from its cache, to the requested $PS_q$ through the optimal path found by the $SM_{TR}$ and the same is intimated to the requested $PS_q$ and hence the client waiting time is relatively higher, but acceptable with high *QoS*.

If the $V_i$ is not present in any of the *PSs* in that $L_p$, then the $IM_{TR}$ Passes the request to the tracker of $NBR(L_p)$. Then the $VDM(NBR(L_p))$ checks its database using perfect hashing, to see whether the $V_i$ is present in any of the *PSs* of its $L_p$. If it is present in one or more PSs, then the $SM(NBR(L_p))$ selects the optimal streaming path from the selected $PS(NBR(L_p))$ to the requested $PS_q$ and intimates the same to $IM(L_p)$. Then the $SM(L_p)$ in turn initiates the streaming of $V_i$ to the requested $PS_q$ through the optimal path, and the same is intimated to the requested $PS_q$ and hence client waiting time is comparatively high but acceptable because it bypasses the downloading of the complete video from *MMS* using *MMS-PS WAN bandwidth*.

If the $V_i$ is not present in any of the *PSs* of its $NBR(L_p)$ also, then the $TR(L_p)$ modules decides to download the $V_i$ from *MMS* to $PS_q$. So the $IM_{TR}$ coordinates with *MMS* to download the $V_i$, and hence the waiting time is very high, but the probability of downloading the complete video from *MMS* is very less as shown by our simulation results.

Whenever the sufficient buffer and bandwidth is not available in the above operation the user request is rejected.

## 5. Simulation Model

In our simulation model we have a single *MMS* and a group of 6 *TRs*. All these *TRs* are interconnected among themselves in a ring fashion. Each of these *TR* is in turn connected to a set of 6 *PSs*. These *PSs* are again interconnected among themselves in a ring fashion. To each of this PS, 25 clients are connected. We use the video hit ratio (*VHR*), the average client waiting time $y$ and network usage as parameters to measure the performance of our proposed approach more correctly by comparing the results of RPPCL, GWQ and PRLS algorithms. In addition we also use the *WAN* bandwidth usage on *MMS-PS* path and probability of accessing the main server as the performance metrics.

Table 1: Simulation Values

| Notation | System Parameters | US Letter Paper |
|---|---|---|
| S | Video Size | 25 to 1120 min |
| $C_{MMS}$ | Cache Size (*MMS*) | 2000blocks |
| $C_{TR}$ | Cache Size(*TR*) | 800(40%) |
| $C_{PS}$ | Cache Size(*PS*) | 300(15%) |
| λ | Mean request arrival rate | 45 reqs/hr |

We assume that the request distribution of the videos follows a zipf-like distribution. The user request rate at each *PS* is 35-50 requests per hour. The ratio of cache sizes at different elements like *MMS*, *TR* and *PS* is set to $C_{MMS} : C_{TR} : C_{PS} = 10:4:2$ and transmission delay between the proxy and the client, proxy to proxy and *TR* to *PS* as 120sec, transmission delay between the main server and the proxy as 480 to 600sec, transmission delay between tracker to tracker 240sec, the size of the cached *[(pref-1)+(pref-2)]* video as 280MB to 1120MB(25-min-1hr) in proportion to its popularity.

## 6. Simulation Results

The simulation results presented below are an average of several simulations conducted on the model

Our main focus was to minimize the client waiting time via exploiting load sharing among the PSs of $L_p$. Fig.9 shows the total number of requests served from the system, the average number of requests served immediately at $PS_q$ as 51%, the average number of requests served from $(L_p+NBR[L_p])$ as 34%, and the average number of requests served from *MMS,* that is only 15% which is very less. The corresponding average waiting time required for serving (pref-1) immediately from PS, from other *PS* of $L_p$ *(Lp+NBR[Lp])* and from *MMS* is shown in the fig. 5.

As the *(pref-1)* of most frequently asked videos have been cached and streamed from the $PS_q$ of $L_p$ and $NBR[L_p]$, with the cooperation of various modules of *PSs,* and the coordination of modules of *TR* of $L_p$, Our scheme has achieved a very high video hit ratio ( 86%) as shown in Fig 6. So the local and global popularity based replication of mostly accessed videos at the respective

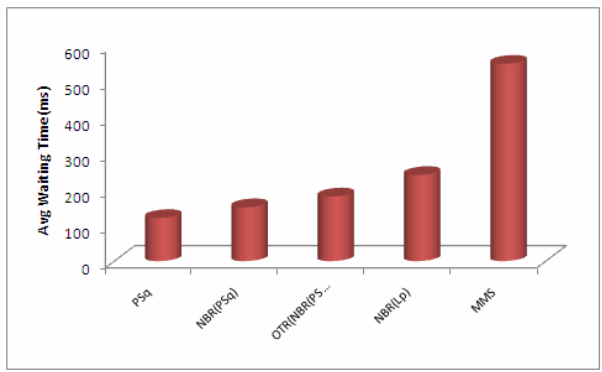

Figure 5 Avg waiting time for videos from PSq, NBR[PSq],Lp,NBR[Lp] and from MMS

*PSs* of *LPSG* has significantly reduced the waiting time for the user when compared to GWQ and PRLS as shown in Fig.6 and Fig.7.

Thus more (80% - 86% of the video) number of blocks of requested videos are cached and streamed from $L_p$, by sharing the videos among the proxies and *TR* of $L_p$. So when there is a request for any of these $i^{th}$ video, streaming starts from one of the *PS* immediately and hence *client waiting time*, *network usage* from MMS to proxy is very less as shown in fig. 6 and 8, and in turn transmission cost, transmission time is also reduced. GWQ also reduces the waiting time by balancing the load between heavily loaded and lightly loaded proxy servers. But it still introduces the unnecessary network traffic flows from remote servers.

If the requested videos are present at $NBR(PS_q)$ of $L_p$, then these videos are streamed from $NBR(PS_q)$ to the client through $PS_q$, so the waiting time for these videos is very small. If the requested videos are present in $L_p$-$NBR(PS_q)$, then these videos are streamed from $L_p$- $NBR(PS_q)$ to the client through $PS_q$, so the waiting time for these videos is relatively higher, Otherwise also, some good number of videos are served from $NBR(L_p)$, which reduces frequent downloading of requested videos from *MMS* to the $PS_q$ which in turn reduces the initial play out delay for the clients for the requested videos which are not which are not present at $PS_q$ as shown in Fig.5 and 9.

*MMS* has been contacted for very few (15-25% of the videos) number of videos, when the $V_i$ is neither present in that $L_p$, nor in $NBR(L_p)$. Even though the initial startup delay and transmission cost seems to be more it is acceptable because on an average *(pref-1)* and *(pref-2)* of

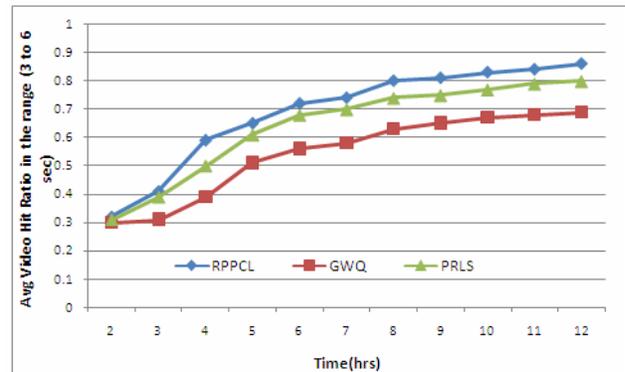

Figure 6 Avg Video Hit Ratio with RPPCL, GWQ and PRLS algorithms

nearly 85% of the videos are cached and streamed from $L_p$ and $NBR(L_p)$ by assuring high *QoS* as shown in Fig.9

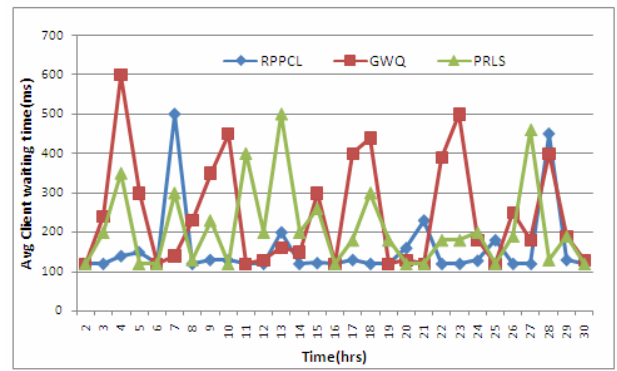

Figure 7 Avg waiting time using RPPCL,GWQ and PRLS algorithms

and only about 15% of the videos are downloaded from *MMS* which has drastically reduced the client waiting time. Hence our proposed approach has successfully achieved the load balance among the interconnected *PSs* of $L_p$ and $[NBR(L_p)]$.

## 7. Conclusions

In this paper we have proposed an efficient regional (local and global) popularity based replication, prefix caching and load sharing (RPPCL) algorithm with the architecture. In which all *PSs* cooperate with each other to achieve reduced *client waiting time* and increased *video hit ratio*, by caching (replicating) and streaming maximum portion ((pref-1)+(pref+2)) of most frequently requested videos among the proxies of $L_p$. Our simulation results demonstrated that our proposed approach has reduced the *client waiting time* for the videos requested at $PS_q$, *average network traffic* of the system, and also the load of *MMS* by the regional popularity based replication of most popular videos at appropriate *PSs* of $L_p$. And sharing of these videos among the proxies of the system with the coordination of Tracker also reduces the server-to-client network usage, transmission cost and time, maintains high *QoS* for the users when compare to GWQ and PRLS algorithms. The future work is being carried out to improve the performance of the system by writing an algorithm to handling the failure of the coordinator tracker.

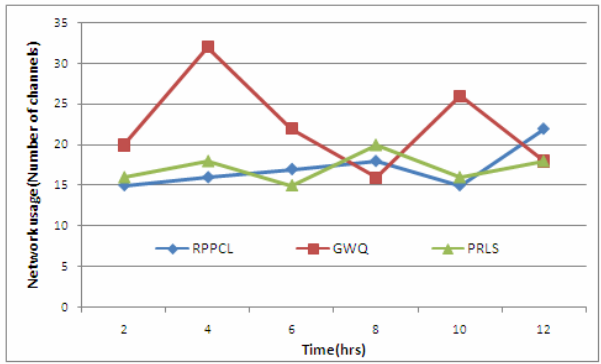

Figure 8 Network bandwidth usage by RPPCL, GWQ and PRLS algorithms

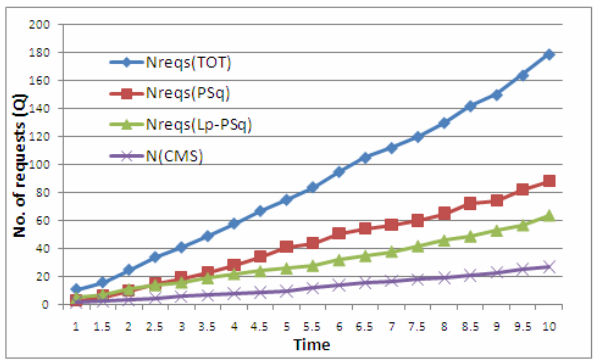

Fig. 9 Total No. of requests served from Lp, PSq, (Lp-PSq), CMS

## References

[1] Bing Wang, Subhabrata Sen, Micah Adler and don Towsley "Optmal Proxy cache Alloction for Efficient Streaming Media Distribution" IEEE Transaction on multimedia, vol. 6, No. 2, April 2004.

[2] H.S.Guruprasad, M Dakshayini et. el "Dynamic Buffer Allocation for VoD System Based on Popularity" proceedings of NCIICT 2006, PSG College of Technology, Coibatore, 13- 17, WWW. psgtech. edu/ NCIICT/ files/ NCIICT06.

[3] Y.C Tay and HweeHwa Pang, "Load Sharing in Distributed Multimedia-On-Demand Systems", *IEEE Transactions on Knowledge and data Engineering*, Vol.12, No.3, May/June 2000.

[4] S. González, A. Navarro, J. López and E.L. Zapata, "Load Sharing in Distributed VoD Systems", *Int'l Conf. on Advances in Infrastructure for e- Education, e-Science, and e-Medicine on the Internet (SSGRR 2002w)*, L'Aquila, Italy, January 21-27, 2002.

[5] Yuewei Wang, Zhi-Li Zhang, David H.C. Du, and Dongli Su "A Network-Conscious Approach to End-to-End Video Delivery over Wide Area Networks Using Proxy Servers", *IEEE INFOCOM*, pp 660-667, April 1998.

[6] Alan T.S lp, Jiangchuan Liu, John C.S.Lui COPACC: An Architectureof cooperative proxy-client caching System for On-Demand Media Streaming. A technical report.

[7] S.-H. Gary Chan, Fouad Tobagi, "distributed Servers Architecture for Networked Video Services", IEEE. Transactions on networking, vol.9, No. 2, Aprol.

[8] S. Acharya and B. C. Smith, "Middleman: A video caching proxy server". in *Proc. of NOSSDAV*, June 2000.

[9] A. Feldmann, R. Caceres, "Performance of Web Proxy Caching in Heterogeneous Bandwidth Environments", In Proc. Of IEEE INFOCOM '99, March 1999.

[10] P. A. Chou and Z. Miao. Rate-distortion optimized streaming of packetized media. Technical Report MSR-TR-2001-35, Microsoft Research Center, February 2001.

[11] Dr.Mahmood Ashraf Khan, Prf.Go-Hasegawa, Yoshiaki Taniguchi "QoS Multimedia Network Architecture" murata Laboratory, Osaka University, Japan.

[12] Lian Shen, Wei Tu, and Eckehard Steinbach "A Flexible Starting Point based Partial Caching Algorithm For Video On Demand", 1-4244-1017-7/07@2007 IEEE.

[12] B. Wang, S. Sen, M. adler, and D. Towsley, " Optmal Proxy Cache Allocation For Efficient Streaming Media Distribution," In IEEE INFOCOM, june 2002.IEEE.

**Dr.T R Gopalakrishnan Nair** holds M.Tech. (IISc, Bangalore) and Ph.D. degree in Computer Science. He has 3 decades experience in Computer Science and Engineering through research, industry and education. He has published several papers and holds patents in multi domains. He won the PARAM Award for technology innovation. Currently he is the Director of Research and Industry in Dayananda Sagar Institutions, Bangalore, India.

**M Dakshayini.** holds M.Tech (VTU Belgaum) in computer science securing second rank . She has one and a half decades experience in teaching field. She has published many papers. Currently she is working as a teaching faculty in the department of Information science and engineering at BMS College Of Engineering, Bangalore ,India.